# Large energy storage efficiency of the dielectric layer of graphene nanocapacitors

A. Bezryadin[1], A. Belkin[1], E. Ilin[2], M. Pak[3], Eugene V. Colla[1], and A. Hubler[1]

[1] Department of Physics, University of Illinois at Urbana-Champaign, Urbana, IL 61801, USA

[2] Department of Physics, Far-Eastern Federal University, Vladivostok, Russia

[3] Department of Chemistry, University of Illinois at Urbana-Champaign, Urbana, IL 61801, USA

**ABSTRACT.** *Electric capacitors are commonly used in electronic circuits for short-term storage of small amounts of energy. It is desirable however to use capacitors to store much larger energy amounts to replace rechargeable batteries. Unfortunately, the existing capacitors cannot store a sufficient energy to be able to replace common electrochemical energy storage systems. Here we examine energy storage capabilities of graphene nanocapacitors, which are try-layers involving an Al film, $Al_2O_3$ dielectric layer, and a single layer of carbon atoms, i.e., graphene. This is a purely electronic capacitor and therefore it can function in a wide temperature interval. The capacitor shows a high dielectric breakdown electric field strength, of the order of 1000 kV/mm (i.e., 1GV/m), which is much larger than the table value of the $Al_2O_3$ dielectric strength. The corresponding energy density is 10 to 100 times larger than the energy density of a common electrolytic capacitor. Moreover, we discover that the amount of charge stored in the dielectric layer can be equal or can even exceed the amount of charge stored on the capacitor plates. The dielectric discharge current follows a power-law time dependence. We suggest a model to explain this behavior.*

## INTRODUCTION

Conventional electric capacitors are typically considered to be not very promising candidates for energy storage applications, because the charge stored on the metal plates is limited by the breakdown threshold in the electric field strength, thus limiting the energy density to at most hundreds of J/kg. The use of capacitors as energy storage systems has received a substantial revival of interest recently, due to the possibility of large increases in energy density for capacitors with gap spacing between the plates on a nanometer scale [1,2]. The improvement is expected because in vacuum capacitors the field enhancement factor, $\beta$, characterizing the local enhancement of the average electric field due to surface defects, decreases with the gap [3,4]. Hence, capacitors with the thickness of the insulating layer of the order of a few nanometers, called nanocapacitors, might be able to withstand much higher electric fields before a breakdown damaging the dielectric takes place. Thus, it is expected that the stored energy density can be improved, especially if advanced, light-weight materials, such as graphene, are used to make the plates of the capacitor [5]. Graphene is a smooth material so the field enhancement factor should be low. Graphene has been used previously for making capacitors and other nanoscale devices [6,7,8,9,10].

Nanocapacitors based on graphene or similar metallic nanomaterials capping very thin dielectrics in between have been a subject of active study. A theoretical study of quantum size effects between graphene electrodes and h-BN dielectric was reported for vertical stacking [11] as well as for in plane geometry [12]. Such capacitors have been studied experimentally also by [13,14,15,16].

For practical applications, it is conventionally assumed that most of the electric charge in a standard metal/insulator capacitor is stored on the metal plates. However, in a fully charged capacitor there is usually some extra charge stored in the dielectric layer separating the plates. The dielectric volume charge is usually neglected since it is typically small, of the order of 1 % of the total charge accumulated in the capacitor. Such charge penetration into the dielectric is driven by the strong electric field occurring in the dielectric spacer of a nanocapacitor. Charging of the dielectric layer has been studied previously [17,18,19] but not on graphene-based nanocapacitors. Generally, there are several competing mechanisms by which the energy can be retained in the dielectric layer. Two of them: hindered dipole rotation and displacement of electrons and/or ions in the dielectric, do not involve any transfer of charge between the metal plates and the dielectric [18,19]. Additionally, there is always the possibility of charge penetration into the dielectric due to the movement of hopping charges, i.e. the tunneling of electrons between the metal plates and some localized charge traps in dielectric [20,21,22,]. Moreover, even if new charges do not enter, the charges located on the traps or defects inside the dielectric can tunnel from one defect to another and thus shift their positions significantly to cause additional charge accumulation on the plates of the capacitor.

Here we demonstrate that the energy stored in the dielectric layer can be equal and can even exceed the energy stored on the plates of the capacitor. This fact offers a new way to expand capacitor's energy storage capability of nanocapacitors. This phenomenon occurs because the used dielectric layer is very thin, typically about 7 nm of $Al_2O_3$. Thus, electrons can penetrate into it by means of quantum tunneling and therefore can efficiently populate the dielectric. Moreover, it is shown that such type of energy storage device is stable against accidental shortages of the plates of the capacitor. This is because the removal of the charge accumulated by dielectric is a relatively slower process, which is controlled by its internal charge dynamics, generally described by relaxation theory [23]. We find that even if the plates are shorted and the charge is completely removed from them, we can still recover from the dielectric a charge which is about equal to the charge which was present on the plates before the discharge happened. Finally, a capacitance measurements of our graphene nanocapacitors exhibits a five-fold increase of the capacitance of the device at low frequencies. This fact is an indication of the extremely large charge storage on the dielectric layer, which is characterized by a comparatively slow charging and discharging.

Here we combine a nanoscale dielectric layer and a graphene capacitor plate in one device and achieve an enhanced energy density. Based on detailed measurements we conjecture that a large amount of energy can be stored inside the dielectric. Namely, it is estimated that the amount of charge and energy stored in the dielectric layer is about the same as the amount of charge and energy stored on the plates of the capacitor. We present a simple and transparent model of the charge storage which offers a semi-quantitative explanation of the observed dependence of the discharge current, $I_d$, versus time, $t$. The release of the charge stored in the dielectric produces a discharge current which decreases with time as a power-law, $I_d \sim t^{-\alpha}$, where $\alpha \approx 1$, i.e. the current generated by the dielectric layer of the capacitor is inversely proportional to the time of discharge (approximately). On the other hand, the charge stored on the plates is released faster, generating an exponentially decreasing function of time, as expected.

## SAMPLE FABRICATION

Typical nanocapacitors used in this study were made with an 8 to 12 nm thick Al film coated by a 5 to 10 nm film of Al2O3 (the dielectric) and, a single layer of graphene placed on top of the oxide. The oxide was deposited using a commercial atomic layer deposition (ALD) system (Savannah S100 Atomic layer deposition (Cambridge Nanotech)). Graphene layers have been deposited using "*Trivial Transfer Graphene*" kits provided by *ACS Material*.

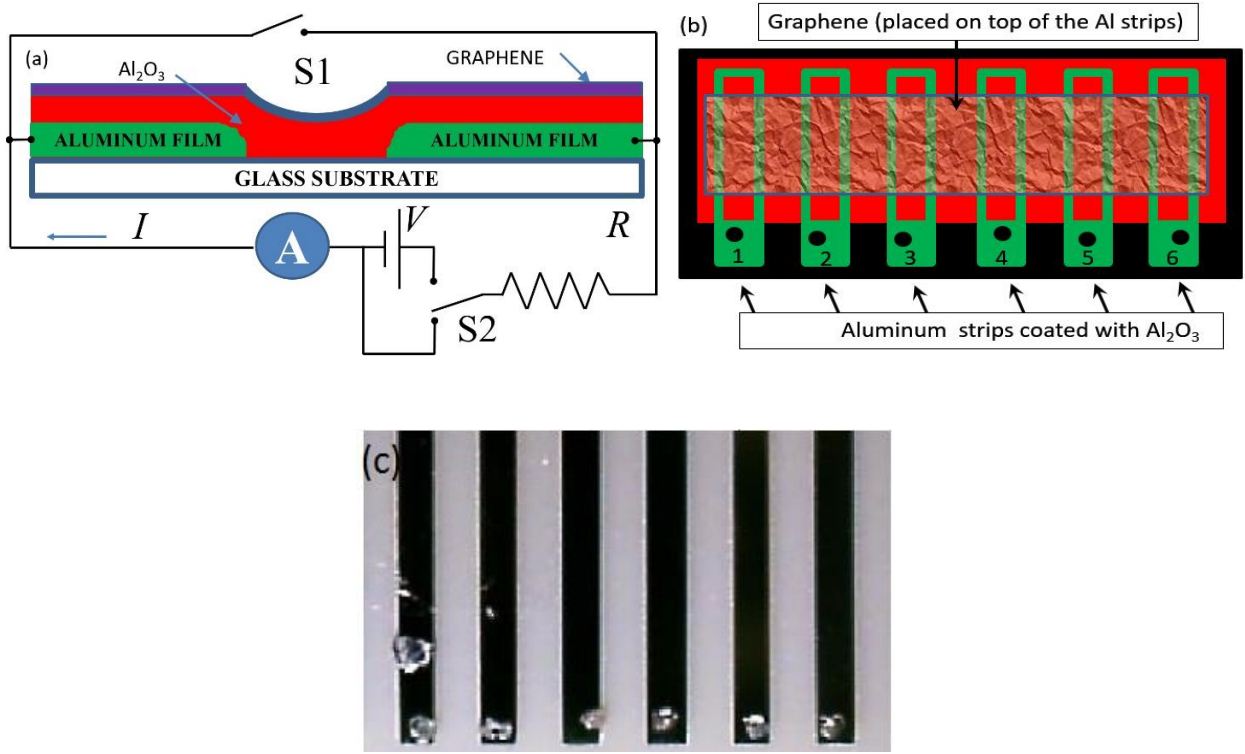

*Figure 1. (a) Schematic of the sample and the measurement scheme. An Al film (~10 nm) (green) is deposited on a glass substrate (white) and covered with alumina (Al₂O₃) (~10 nm) (red), which serves as dielectric layer of the capacitor. The top electrode is a single monolithic layer of graphene (blue). Graphene has metallic conductivity and acts here as the top metallic plate of the capacitors. The width of the Al electrodes is 0.75 mm and the gap between them is 1 mm. The switch S1 is used to instantaneously discharge the capacitor plates. The switch S2 allows to either charge the capacitor through the resistor R and to discharge through the same resistor. (b) Top view of the nano-capacitors. The connections are made to the Al strips (but not to graphene), e.g., #1 and #2. Thus, effectively, the sample consists of two identical capacitors, of the type Al-Al2O3-graphen, connected in series. Possible position of the electrical contacts (indium dots) to the double-capacitor are shown by black circles. (c) Image of the sample. The black strips are the Al electrodes. In dots, used to make contacts to the electrodes, are visible at the bottom. Graphene, which is placed in the central part of the sample, is too thin to be visible in such image.*

The schematic of the sample, image of the sample and the measurement scheme is shown in Fig.1. The sample contains multiple Al electrodes (Fig.1b, c), any pair of which can be used as unique double-capacitor (e.g., 1 and 2). The length of the graphene-covered section of each Al electrode is $L = 5$ or 10 mm and the width of each Al strip is $w$=0.75 mm.

As is clear from Fig.1, the device is not just a single capacitor of the type Al-Al$_2$O$_3$-G, but, rather, it is composed of two nominally identical capacitor connected in series through the graphene film in the middle. This is because, in the current realization, we do not make a contact to graphene, but both contacts are made to two Al strips, say #1 and #2. The capacitor on the left electrode (#1) of the device has a structure Al (green) - Al2O3 (red) – graphene (blue). The capacitor on the next electrode (#2) has the same structure, so the entire sequence is Al-Al2O3-G---G-Al2O3-Al. Such design allows us to avoid making a direct contact to graphene, which is quite fragile. The resulting devices will be called "double-capacitors" since they represent two capacitors connected in series. The gap between the Al electrodes is 1 mm wide and the resistance of the graphene in the middle, between the two capacitors, is of the order of 1 k$\Omega$. Since the capacitance of the sample is of the order of C~10 nF, the fastest discharge time is of the order of 10 $\mu$s, in the case if the leads are short-circuited.

## MEASUREMENTS

### Type I measurement

In the Type I measurement we observe both the charge on the plates of the capacitor and the charge stored in the dielectric, in the same measurement run. In this case the switch S1 is always open (i.e., disconnected). First, we charge the capacitor up to some voltage $V$, typically $V$=3V. The voltage is supplied by Keithley 6517; the same device is used to measure the current ("A"). To charge the capacitor the switch S2 is connected upward (Fig.1a). After the capacitor is fully charged, the switch S2 is connected down (so that the voltage source is excluded from the circuit) and the discharge current is measured as a function of time (Fig.2; red curve). The discharging rates is limited by the resistor $R$. If $R$ is sufficiently large (e.g., 100 M$\Omega$ or 1 G$\Omega$), the discharge takes many seconds and can be easily measured using Keithley 6571. Thus, the type I measurement produces results as shown in Fig.2. This type of measurement, as will be discussed in detail later, shows two distinct stages of the discharging process. The first stage is the exponential discharge of the capacitor plates. The second stage is the discharge of the charges stored in the dielectric film separating the plates of the capacitor. This is a slower process, controlled by the internal charge dynamics in the insulator. The discharge current obeys a power law, as is illustrated by the linear dashed fitting line (Fig.2). Note that a power-law dependence appears linear in the log-log axis format.

## Type II measurement

The second type of the measurement algorithm is used when we want to quantify, exclusively, the charge stored in the dielectric, $Q_d$. For this purpose, we first charge the capacitor fully. For this purpose, we open the switch S1 and connect S2 upward, and apply a voltage $V$ for a time ~10 min (Fig.1a). When the charging is complete, the switch S2 is turned downward, to exclude the charging voltage source. Then, immediately after, the switch S1 is used to short-circuit the capacitor plates to remove all charge stored on the capacitor plates (i.e., Al strips and graphene film). As the switch S1 is on, the charge stored on the plates of the capacitor discharges essentially instantaneously (within ~ 10 microseconds). After the capacitor is discharged the switch S1 is set "off" again, i.e., the path short-circuiting the plates is removed. At this point the switch S2 is still connected downward, thus enabling a discharge circuit involving the capacitor, the ammeter, and the resistor, all connected in series. Thus, the current flowing between the plates of the studied capacitor is measured with the ammeter. If the charge is not stored in the dielectric but only on the plates then the current in such measurement algorithm would be exactly zero, since the plates are fully discharged at time $t$=0, when the measurement begins. Yet, as we will discuss below (Fig.6), the measured current and the corresponding integrated charge are both quite large. In fact, the charge stored on the dielectric is approximately equal to or even sometimes higher than the charge stored on the plates.

## RESULTS

First, we perform Type I measurements. The capacitor is charged from a voltage source at $V$=3V. Then the capacitor is allowed to discharge through the resistor $R$ and the current $I$ is measured. Typical results are shown in Fig.2. Each such charging process exhibits two distinct stages. First, we observe the initial exponential discharge representing the charge on the plates of the capacitor. The second stage is a power-law dependence of the current at time scale longer than ~10 s, which represent the charge stored in the dielectric layer of the capacitor (Fig.2).

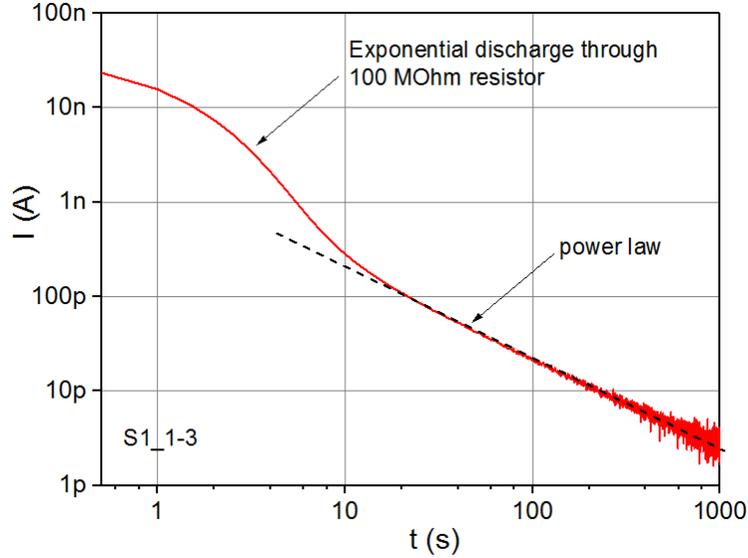

*Figure 2. Sample S1_1-3. The discharge current is shown versus time, in log-log format. The discharge is done after the capacitor was charged to 3 V. the discharge current flows through the 100 MΩ resistor and an ammeter (Fig.1a). Initially, within first 10 seconds or so, the discharge current drops exponentially, as expected for an ideal capacitor shunted by a resistor. At larger time the current follows a power law I~1/t. This process is associated with the drainage of the charge stored in the dielectric layer of the capacitor.*

The initial regime is the exponential dependence, as is expected for a discharge process of a simple parallel-plate ideal capacitors. A numerical fitting shows that the dependence is well described by an exponent, initially, as is confirmed by the fits in Fig.3 (dashed blue curve). There the large black circles represent a discharge process through the resistor $R$. The time scale $t$-$t_2$ corresponds to the time elapsed from the beginning of the discharge process. The time interval $t_2$ is the delay between the end of the charging process and the beginning of the discharging. During the time interval $t_2$ the sample is completely disconnected from any circuit (this is done to see how reliable the charge storage is on this type of capacitor). The fit is shown by the blue dashed line and corresponds to the expected exponential decrease of the current as $I(t)$=$V(t)$/R=$Q(t)/RC$=[$V(t_0)/R$] × exp[-$(t$-$t_0)/\tau$], where $\tau$=$RC$ is the time constant of the circuit including the resistor. Since the results represent a delayed measurement, the initial voltage is lower than the applied voltage, i.e., $V(t_0) < V(0) = 3$ V.

The fit (blue dashed line) agrees well with the data (black circles), up to the point where the current drops by a factor about 20 compared to its initial value, which corresponds to the discharge duration $t$-$t_2$ ~3s. This initial stage represents the discharging of the capacitor plates, the rate of which is limited by the resistor *R*. At larger times we observe current significantly exceeding the value expected from the exponentially decreasing fitting function. Moreover, at larger times, i.e. $t$-$t_2$ >3s, the current-versus-time dependence is approximately linear in the log-log plot. This fact suggests that at larger times a different mechanism of the energy storage is at play. This mechanism, as we will discuss in detail below, represents the charge stored inside the dielectric, and not on the plates of the capacitor. In the subsequent sections we will show that this charge is very significant and comparable to the charge stored on the capacitor plates.

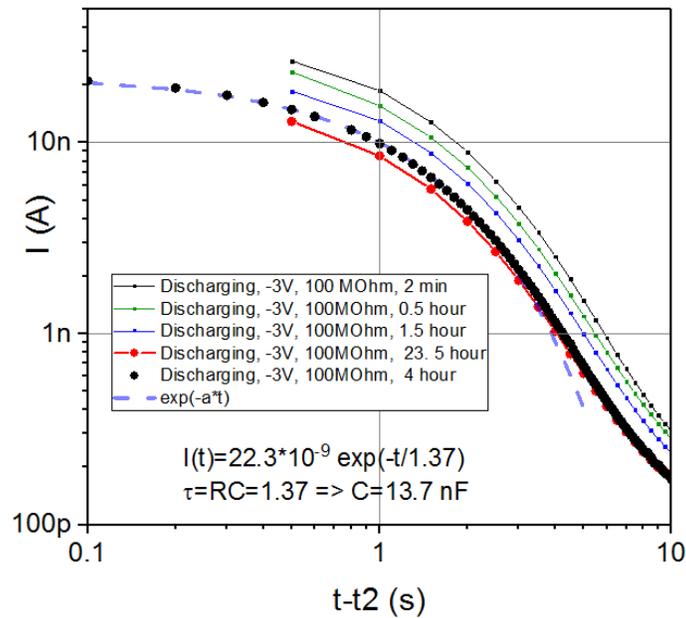

*Figure 3. Sample S1_1-3. The discharge process of a double-nano-capacitor (see Fig.1), C, through a resistor R=100 MΩ. All the curves represent delayed discharge measurements: After the charging was completed, the capacitor was disconnected from all leads for a time interval $t_2$. This is done to test the internal leakage of the capacitor. At time $t_2$ the capacitor is connected to the ammeter through the resistor R and the discharge curve is measured. The delay interval for the curves presented, from the top to the bottom, was $t_2$=2 min, 0.5 h, 1.5 h, 4 h, and 23.5 h. At t=$t_2$ the capacitor was connected to the ammeter through the resistor R and the current was measured versus time. The measurements are shown by symbols which are connected by lines, except the curve corresponding to $t_2$=4 h, in which case the measurement is shown by black circles. The exponential time-dependent fit (blue dashed curve) is shown for the case of $t_2$=4 h. The initial current was 22.3 nA, according to the fit, which corresponds to the initial voltage of 2.23 V (after a 4 h waiting period with disconnected leads).*

The initial exponential fit allows us to determine the capacitance of the sample. The fitting function and the fitting parameters are shown in Fig.3. The time constant is $\tau=RC$=1.37s. Since $R$=$10^8$ $\Omega$, one gets $C$=13.7 nF. Let us compare this value to the theoretical value, $C$=$(1/2)\varepsilon\varepsilon_0 Lw/d$, expected from the sample geometry. Here $\varepsilon_0$ is the permittivity of vacuum, $\varepsilon\sim9$ is dielectric constant of aluminum oxide. For the sample of Fig.3 (sample S1_1-3), the length of the electrodes is $L\sim10$ mm and the width is $w\sim0.75$ mm. The thickness of the dielectric ($Al_2O_3$) is $d\sim10$ nm. The factor 1/2 is included because in our samples two nominally identical capacitors are connected in series. Thus, we get $C\sim30$ nF. This estimate is somewhat larger than the measured value. We ascribe the difference to the expectation that the dielectric constant of the aluminum oxide is reduced at nanoscale, due to disorder, since our dielectric films are amorphous.

## INTERNAL LEAKAGE

For energy storage applications it is important that the capacitor holds charge for extended periods of time. The main reason why the energy can be lost is the internal leakage between the plates of the capacitor, through the layer of the dielectric. Standard value for the resistivity of $Al_2O_3$ is $10^{14}$ $\Omega$ or larger. For our geometry (sample S1_1-3), the corresponding resistance of the capacitor dielectric layer is $R_C\sim10^{11}$ $\Omega$, and the corresponding internal time constant would be $\tau_i=R_C C\sim30$ min. We will see below that the actual measured internal discharge times are even better than this estimate.

Let us compare this estimate to the experimental values, which can be found from the data of Fig.3. The black circles represent a discharge current-versus-time curve measured after a delay of $t_2$=4 hours. In this experiment, the capacitor was charged to 3 V, then disconnected from circuit completely for 4 hours. During this disconnection time, only the internal leakage was contributing to the capacitor discharge. Then, after 4 hours, the sample was reconnected to the measurement circuit (Fig.1) and the discharge curve was measured with a series resistor of 100 M$\Omega$. The fit (Fig.3, blue dashed curve) gives us the value of the current at the beginning of the discharge curve measurement, which was $I(t_2)$ =22.3 nA. Thus, the voltage on the capacitor plates, after the waiting period of 4 hours, was $V(t_2)=RI(t_2)$=22.3nA*100 M$\Omega$=2.23V. Since the initial voltage was $V(0)$=3V, the internal leakage time $\tau_i$ is estimated as $\tau_i=t_2/\ln[V(0)/V(t_2)]$=13 h. This is much better than the above estimate, suggesting that the effective resistivity of our nanometer-thick $Al_2O_3$ dielectric is about $10^{15}$ $\Omega$ or somewhat higher, which indicates an excellent quality of our ALD oxide.

Another, similar way to estimate the leakage of the charge through the dielectric is to compare the discharge curves measured after $t_2$=4 h waiting period and $t_2$=23.5 h waiting period (red circles), plotted in Fig.3. Each discharge curve is $V(t)=V_0(t_2) \times \exp[-(t-t_2)/\tau]$, where $\tau=RC$ is the discharge time defined by the series resistor $R$. And, as was discussed above, the internal leakage is described by its own exponent $V_0(t_2)=V_0(0) \times \exp[-t_2/\tau_i]$, where $t_2$ is the duration when the capacitor is disconnected from the circuit completely. In other words, $t_2$ is the time interval between the moment when the capacitor charging is finished and the time when its discharge is started by connecting it to the resistor and the ammeter. The discharge through the dielectric layer will be called internal discharge. The discharge through the resistor will be called controlled discharge. Obviously, the controlled discharge goes much faster since the resistor value is much lower than the leakage resistance of the dielectric layer.

The controlled discharge current is $I_b(t-t_{2b})=[V_0(t_{2b})/R] \times \exp[-(t-t_{2b})/\tau]$ for the black-circle curve and $I_r(t-t_{2r})=[V_0(t_{2r})/R] \times \exp[-(t-t_{2r})/\tau]$ for the red-circle curve. Here $t_{2b}$=4 h and $t_{2r}$=23.5 h. The compare the black-circle and the red-circle measurements in Fig.3 we choose the same controlled discharge duration on both curves, namely $t-t_{2r}= t-t_{2b}$ =0.5 s. This time is chosen because the current at this time is relatively high and also because the corresponding data point is available on both curves. The ratio of the controlled discharge currents is $I_b(t-t_{2b})/I_r(t-t_{2r})=[V_0(t_{2b})/V_0(t_{2r})] \times \{\exp[-(t-t_{2r})/\tau]/\exp[-(t-t_{2b})/\tau\}=V_0(t_{2b})/V_0(t_{2r})$. It is assumed that the voltage at the beginning of the controlled discharge is reduced due to the preceding spontaneous discharge according to the usual discharge formula $V_0(t_{2b})=V_0(0) \times \exp[-t_{2b}/\tau_i]$ and $V_0(t_{2r})=V_0(0) \times \exp[-t_{2r}/\tau_i]$. So $V_0(t_{2b})/V_0(t_{2r})= \exp[(t_{2r}-t_{2b})/\tau_i]$ and $I_b(t-t_{2b})/I_r(t-t_{2r})=\exp[(t_{2r}-t_{2b})/\tau_i]$. Finally, we get a relationship between the internal discharge time constant and the initial current in the controlled discharge measurement as follows $\tau_i=(t_{2r}-t_{2b})/\ln(I_b(t-t_{2b})/I_r(t-t_{2r}))$. Thus, we get the internal discharge time constant $\tau_i$=135h. The results suggest that the capacitor can be useful to store charge for many days. This time is larger than the independent estimate given above because the discharge might go faster when the voltage is higher.

# BULK CHARGE MEASUREMENTS

The most significant result of this work is the observation of a large additional charge, $Q_d$, stored in the dielectric layer of the capacitor. We find that the charge $Q_d$ accumulated by the layer of Aluminum oxide is approximately equal, and sometimes even larger than, the charge $Q_p=CV$, stored on the plates of the capacitor. Here $C$ is the geometric capacitance. It is possible to measure these two charges, $Q_p$ and $Q_d$, separately since the discharge of the capacitor plates is an exponential function of time, with the time constant controlled by the resistor $R$ shunting the capacitor. The discharge of the dielectric-stored charge $Q_d$ is a much slower process, usually described by a power-law dependence on time. The dielectric charge release rate is limited by some internal charge migration dynamics inside the dielectric layer.

## *Exponential Fitting Analysis*

Here we present an example of the data analysis intended to quantify the charge stored on the dielectric. To achieve this, we first quantify $Q_p$. The charge on the capacitor metal plates, $Q_p$, is well defined since it follows an exponential dependence on time if the capacitor is charged or discharged through a calibrated resistor. The current flowing to the capacitor plates can be fitted with an exponential function, which can be easily extrapolated to the entire time interval, from zero to infinity, and integrated analytically. The total charge will be estimated by the numerical integration of the total current, as the capacitor is charged. As we will discuss below, the charging current follows the expected exponential function initially, but it exceeds the exponential function at large times.

In this test we apply a bias voltage of 3V to a double-capacitor through a series resistor resister of $R=1$ G$\Omega$ and measure the current as a function of time, $I(t)$ (Fig.4). The schematic of the experiment as is shown in Fig.1, where the switch S1 is open in the test discussed here. The initial charge on the capacitor is zero. At $t=0$ a voltage is applied $V=3$V, which results in a charging current, which decreases with time as the voltage on the capacitor increases. An example of such a measurement is shown in Fig.4. There, the data are shown as black squares and the blue straight line is the best exponential fit corresponding to the initial part of the plot, $I_{fit}=I_0\exp(-t/\tau)$. The fitting constants are $I_0=2.911$ nA and $\tau=10.06$ s. The key observation here is that the charging current decreases slower than exponential, except the initial region of a duration of a few seconds. This fact indicates that, in addition to charging the plates of the capacitor, the dielectric is also being charged. Below we explain how the charge stored on the dielectric can be evaluated numerically. The general approach is to first determine the charge on the plates of the capacitor, $Q_p$, by integrating the best-fit exponent. Then we will integrate the entire $I(t)$ curve and find the total charge $Q_t$. Finally, the charge on the dielectric layer will be naturally calculated as $Q_d=Q_t-Q_p$.

The charge on the plates is defined by a simple integral of the exponent, namely $Q_p=I_0\int\exp(-t/\tau)\mathrm{d}t$. To find the complete charge that would flow into the capacitor plates the integral must be evaluated over the entire time axis, namely from zero to infinity (the charging is assumed to begin at $t=0$). Then the result is $Q_p=I_0\tau=29.3$ nC. Another way to check this result is to first find the corresponding capacitance as $C=\tau/R=10.06$ nF. Then the charge of the capacitor is $Q_p=CV=30.2$ nC, which is similar to the results obtained above.

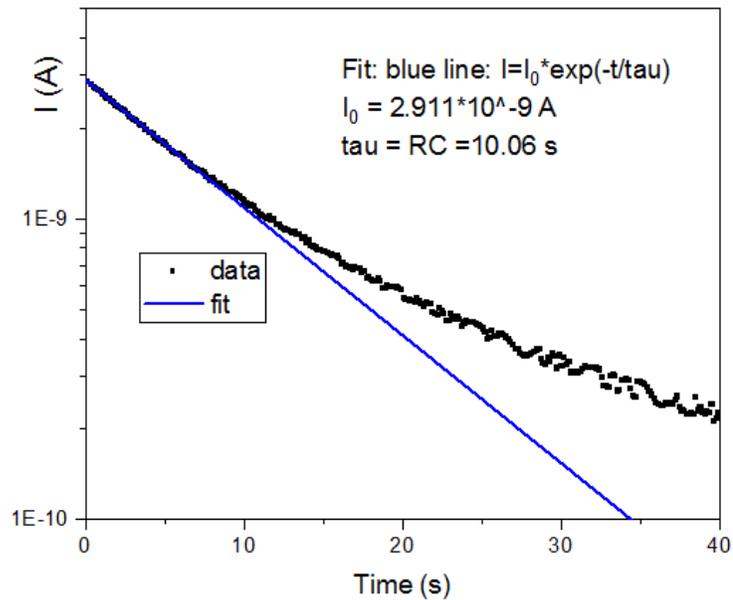

*Figure 4. Sample S1_2-3. The charging current as a function of time for a graphene double-capacitor sample of Fig.1, with series resistor and voltage being R=1GΩ and V=3V. The blue line represents the best exponential fit. The measured current (black squares) appears higher than the expected exponential behavior (blue line), indicating that a significant portion of the current is flowing into the dielectric layer and charging it. The dielectric charge is estimated (see text) to be even higher than the charge localized on the plates of the capacitor.*

The next step of the analysis is the integration of the measured charging $I(t)$ curve. A straightforward numerical integration give the total charge $Q_{t,raw}=132$ nC. We label this estimated charge as "raw" or "uncorrected", because of two reasons. First, the device used to measure current might have an offset of zero. In fact, some offset current is present in any measurement apparatus, due to a finite level of accuracy of any device. A small offset is normally not a big problem, but in the present case, since the charge is evaluated by means of integration, the total charge might appear exaggerated. Second, there could be some leakage current through the insulating layer, when voltage is applied. Again, if the integration is sufficiently long, both the offset and the leakage currents can lead to an overestimated total charge. To compensate for these two effects, we perform a lower bound estimate of the total charge stored. For this purpose we estimate the total charge as $Q_t=\int[I(t)-I_{offset}]dt$. Here the offset current is evaluated as a mean value current at times near the end of the integration interval. In the example of Fig.4 such corrected charge was $Q_t=74.6$ nC, while the estimated current offset was $I_0=13$ pA. If we now take into account the above estimate of the charge on the plates, we can find the charge stored in the dielectric film of the capacitor as $Q_d=Q_t-Q_p=41.3$. Finally, we introduce the efficiency of the dielectric layer to store charge defined as $Eff=Q_d/Q_p=1.4$.

For comparison we have measured a commercial available ceramic capacitor (10 nF, 1kV, Z5U) using the same setup. The observed efficiency of the dielectric storage was determined between $Eff=0.02$ and $Eff=0.025$ (for the bias voltage varying in the range 5-20 V), which is much lower than the value observed on our graphene nanocapacitors.

## Integration Analysis

In this example, the sample has been charged up to 3V and then, the discharge current was measured through a resistor of 100 MΩ.

The current-versus-time dependence has been integrated to produce the charge versus time plot of Fig.5(insert). The most pronounced feature there is the linear increase of the net charge with time, at times larger than ~15000 c. Such linear increase if the capacitor accumulated charge is unphysical and has to be corrected. The explanation is again the offset current. The offset current can be determined as the slope of the linear segment observed at large times (Fig.5, insert). If the offset current is subtracted from the total current, then the integral converges to a constant. This limiting charge is a lower-bound estimate of the total charge stored. There result is given in Fig.5. The total charge is estimated $Q_t=47$ nC. The time constant measured in this experiment was 1.3 s. The resistor was $R=100$ MΩ. Thus, the electric capacitance between the plates is 13 nF. The initial voltage was 2.26 V, and, correspondingly, the charge on the plates was $Q_p=29.4$nC. Therefore, he efficiency in this measurement was found *Eff*=0.6, which is still much larger than has been observed on common ceramic capacitors, for example.

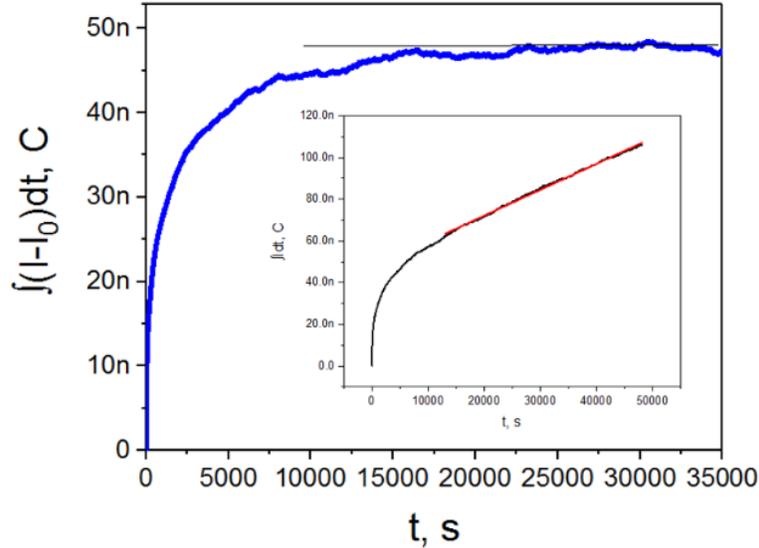

*Figure 5. Insert: The total charge of the capacitor obtained by numerical integration of the corresponding current-versus-time curve measured in a discharge process. The linear segment at large times is due to the offset current. Main figure: Integral of the total current, from which the offset current is subtracted. In this case, the total charge converges to a constant value at large time, which is about 47 nC. This is a lower-bound estimate of the charge stored on the capacitor.*

**Type II measurements, power law dependence, and the dielectric charge storage estimate**

In the experiments above some charge was present on the plates and some charge was present in the dielectric. Thus, disentanglement of these charges had to be carefully addressed. To simplify the analysis and to further confirm our conclusions, here we present a complementary test in which the charge on the capacitor plates is eliminated completely before the measurement begins. The steps of the measurement are as follows: (1) The capacitor is initially charged to a voltage $V$ (1V, or 3V, or 5V) by switching S2 in the downward position. (2) The switch S2 is changed upward. (3) The plates of the capacitor are shorted by closing the switch S1 (Fig.1) for about one second. Note that the estimated discharge time, which the switch S1 is closed, is about 10 microseconds (see above). (4) The switch S1 is opened, so it plays no role any longer. (5) The capacitor is already connected to the ammeter through a series resistor $R$ (Fig.1) since S2 was switched upward is step #2. The discharge current is measured. Such measurement represents the current $I_d$ flowing from the dielectric. Since the plates have been discharged completely before the measurement, therefore $I_\text{p}$=0 and $Q_\text{p}$=0. The results are shown in Fig.6. The plots are presented in log-log format. Here the dots represent the measured points, and the dashed lines are linear fits. The linear fit in the log-log format represents a power-law dependence. The curves measured after the capacitor was charged to 3V and 5V appear to agree with high accuracy with the dependence $I_\text{d}$~1/$t$ (see red and blue data points). The dependence corresponding to the initial charging voltage of 1V appears to follow a power law $I_\text{d}$~$t^{-\alpha}$ with $\alpha$≈1, but, strictly speaking, $\alpha$<1 in this case ($\alpha$ is in fact just slightly less than unity).

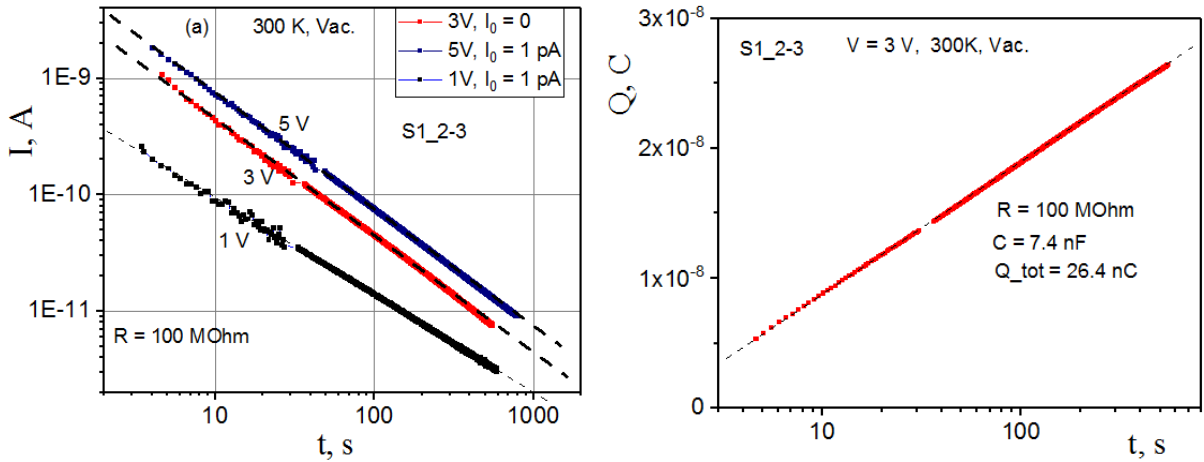

*Figure 6. (a) Sample S1_2-3. The graphene nanocapacitor discharge process, current versus time, shown for three different voltages. The sample was measured in vacuum. Before each measurement the capacitor was charged and then the plates were shorted for one or two seconds to discharge them. Yet the current occurs as an after-effect, which indicates that the removal of the charge on the plates (by connecting them for a short time) does not cause the nanocapacitor to lose all stored energy. In all three cases the discharge current versus time dependence is similar to I(t)~1/t. The exact power of this power law decay can be either slightly smaller or slightly larger than the generic value of -1. For example, the thin dashed line (V=1 V) represents a dependence $t^{-\alpha}$ where $\alpha$ is slightly smaller than 1. The results corresponding to V=3 V and V=5 V correspond to the exact inverse proportionality I(t)=const/t. (b) Integral charge of the discharge curve measure at 3 V.*

It is interesting to note that the integral of a power-law dependence is divergent, either at zero or at infinity or both. Thus, one can expect that an unusually large amount of charge can be stored on the dielectric layer since the current follows a power-law dependence. To make a quantitative conclusion we integrate the current versus time. The result (Fig.6b) is a logarithmic dependence on time. We do not expect any significant contribution of the offset or leakage currents here since the dependence at large times is not linear. The observed logarithmic dependence (which appears linear in the Fig.6b because the x-axis is logarithmic) is exactly what one expects if the current is $I \sim 1/t$, since the charge is the integral of the current over time.

To estimate the efficiency, we use the capacitance measured by a multimeter (LRC meter operating at 1kHz), which turned out to be 7.4 nF. The charging voltage on Fig.6b is 3V. Thus, the charge on the plates is $Q_p = CV = 22.2$ nC. The integrated charge (Fig.6b) is $Q_d = 26.4$ nC. Thus, the efficiency is $Eff = Q_d/Q_p = 1.19$. Thus, in this example analysis, the charge stored on the dielectric appears higher than the charge stored on the plates of the capacitor.

Another conclusion which follows from the results of Fig.6 is that the system is stable against an accidental shortage of the plates of the capacitor. In Fig.6 the plates were shorted before the measurement. Yet, subsequent measurements demonstrate that the energy which remains on the capacitor after it has been shorted is comparable to $CV^2/2$. So, obviously, the energy stored on the dielectric is stable against a short-term short-circuiting of the capacitor plates.

## DISCUSSION

Qualitatively speaking, the accumulation of the charge on the dielectric layer, reported here, resembles the phenomenon of creep [24,25]. In these examples the creep phenomenon describes displacements of molecules. In the case considered here electrons and/or ions, i.e., charged particles might be experiencing creep. Thus, the current obeys a power law, in some general sense analogous to Refs. [26,27]. The logarithmic dependence has been also predicted, for a highly resistive granular system of partially oxidize Al film, in Ref. [26].

Fundamentally, the time-dependent response of a system to a step-function removal of constant stress is most conveniently defined through the Fourier transform of the frequency-dependent response, which is known for a large variety of mechanisms of dielectric absorption. As suggested by Jonscher [24], one can generally distinguish between two fundamentally different types of dielectric response [27]. The first type, termed 'dipolar', is a generalization of the Debye concept of polarization due to the hindered movement of ideal dipoles. This type of polarization leads to zero residual polarization after discharge, with the time dependence of the charge described by the exponential law $\sim exp(-\omega_p t)$, where $\omega_p$ is the loss peak frequency of the dielectric response in the frequency domain. The second type of response describes the behavior of a system where polarization is dominated by slow movement of hopping charge carriers, which is relevant for the charge accumulation due to tunneling of electrons to localized states in dielectric. This type of dielectric response has a fractional power law time dependence $\sim t^{-\alpha}$, and is characterized by finite residual polarization of the system.

In practical applications, the measured property is usually not the residual polarization or the residual charge, but the discharge current. Generally, for times much longer than the $RC$ relaxation time of the capacitor, the discharge current due to different mechanisms of charge retention by dielectric can be described by the fractional power law $\sim t^{-\alpha}$, with different mechanisms resulting in different values of the power $\alpha$ [6]. The varying time dependence of the discharge currents resulting from different mechanisms of charge absorption can in principle be used to eliminate some mechanisms for a particular dielectric, or to analyze the relative contributions of those mechanisms. Additionally, beside the different time dependences, some mechanisms are characterized by dissimilarity between the charging and discharging processes, which can be further used for the same purpose. Specifically, the trapped volume charge formed through charge injection, possibly by tunneling [28], is absorbed and released on different time scales, making its contribution apparent if a large hysteresis is observed between the charge and discharge currents.

Now we estimate the energy density using the example of the sample S1_1-3, the parameters of which are given below Fig.3. The total energy is $E=CV^2/2\sim60$ nJ. If one takes into account the energy stored in the dielectric, then $E\sim100$ nJ. The energy density is $E/(Lwd)\sim1.3$ MJ/m$^3$. This can be converted to energy per unit mass. The density of Al$_2$O$_3$ is 4 g/cm$^3$=4000kg/m$^3$. Thus, the mass of the dielectric layer is 0.3 nano-gramm. So, the energy density (per unit weight) is $\sim330$ J/kg. This is about an order of magnitude better than ordinary electrostatic capacitors. Here we accepted an optimistic point of view assuming that both of the capacitor electrodes can be made of graphene, the weight of which is negligible compared to the weight of the dielectric film. Such nano-capacitors can be used for local energy storage, at cryogenic temperatures for examples, where the ionic and electrolytic capacitors might not work.

## Capacitance versus frequency

The results presented above suggest that the charge stored in the dielectric is large, but it takes a longer time to extract it. Thus, if the capacitance is measured at a sufficiently low frequency, its capacitance should be significantly large than the capacitance measured at a higher frequency. We have tested this hypothesis using a commercial capacitance measurement setup. For measurement of the capacitance was used LCR HP4776 (100Hz – 20 kHz) and lock-in amplifier SR 830 for frequencies below 100Hz. The results are shown in Fig.7, for two similar samples.

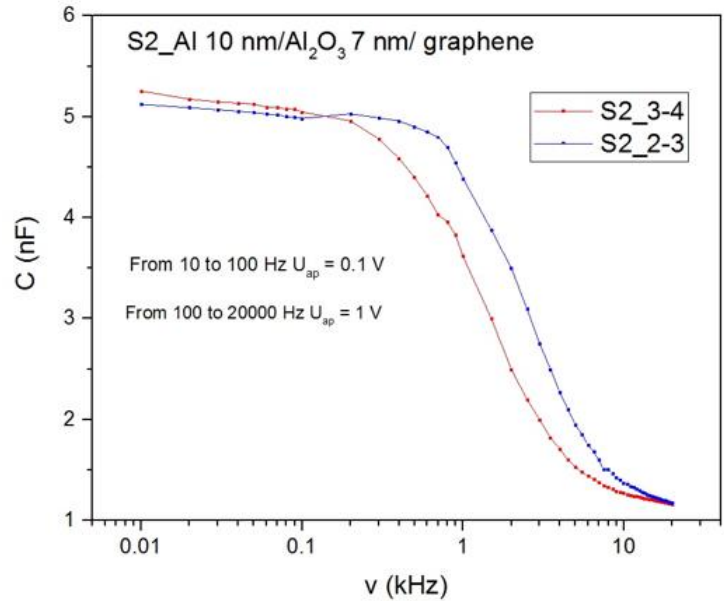

*Figure 7. (a) Capacitance of a double-nano-capacitor with a graphene top electrode measured versus frequency. A roughly five-fold increase of the capacitance is apparent as the frequency is reduced from ~10 kHz to ~10 Hz. The sample design is given in Fig.1.*

The results show that the capacitance remains roughly constant in the interval 10-20 kHz. As the probe frequency is reduced from 10 kHz to 10 Hz, the effective capacitance of the sample increased by a factor five or so. The rapid growth stops near 1 kHz. At lower frequencies the capacitance continues to grow, but significantly slower, and, apparently, linearly, in the log-linear format of Fig.7. This linear increase, at frequencies lower than about 200 Hz, represents a logarithmic growth because the x-axis scale is logarithmic. The reported large storage capacity of the dielectric is related to the logarithmic increase of the capacitance at low frequencies (<200 Hz). Indeed, if $Q$~$\ln(t)$~$\ln(1/f)$ then $C=Q/V=C_0+C_1\ln(1/f)=C_0-C_1\ln(f)$, exactly as observed.

## MODEL

Quantum tunneling has been discussed previously as a mechanism allowing charge penetration into the dielectric layer [29]. Here we propose a simpler and more intuitive model, based on quantum tunneling, which assume many impurities (rather than just one, as discussed in the reference above). As we will see, our model reproduces the power law dependence observed in the discharge current, $I$(t)~$t^{-\alpha}$, with α~1.

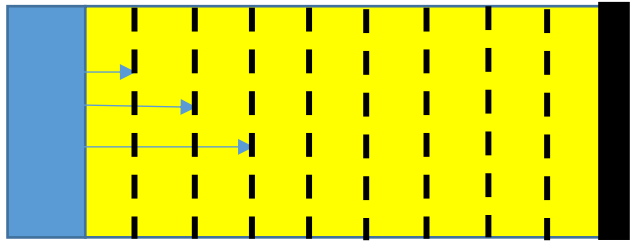

*Figure 8. Proposed model. The capacitor is formed between a metallic Al film (blue) and a layer of graphene (black). The dielectric spacer between the plates (yellow) is $Al_2O_3$. The dashed lines show hypothetical charge traps. They are positioned along straight lines for the sake of an easier analysis. In reality the position of the charge traps is random. We estimate that the number of traps is of the order of one percent of the total number of atoms in the dielectric layer. The distance between the dashed lines is $d_p$.*

The model is illustrated in Fig.8. A plane capacitor is formed between two metallic plates (blue and black) and a dielectric spacer (yellow). The position of the charge traps is shown by dashed lines, schematically. For simplicity of the argument we assume that the charge traps are positioned along the dashed lines, their distance from the left metallic plate being $x_n=d_p \times n$, where $n$=1,2,3.. is an integer, and $d_p$ is the distance between trap planes. Let us assume that the left metallic plate is negatively biased, and the electrons propagate to the right, into the dielectric, by means of quantum tunneling [16] (see the horizontal blue arrows). Note that the match between the energy of the incoming electron and the energy of the state localized in a trap does not have to be perfect since electrons exchange some energy with the thermal bath. All reported experiments are done at room temperature, so thermal fluctuations should help tunneling between states with slightly different energies. The tunneling events can happen from the metallic plate to a trap inside the dielectric; they can also happen inside the dielectric, from one trap to a neighbor trap, but this possibility is not analyzed here. The probability of a tunneling event is controlled only by the distance over which the tunneling occurs, and not by the initial position of the electron.

According to the elementary theory of quantum tunneling, the tunneling probability to the first trap plane (top arrow, Fig.8) can be estimated as $p_1=C\times\exp[-x_1/\xi]=C\times\exp[-d_p/\xi]$. Here $\xi$ is the localization length of the electronic wave function. The tunneling rate to the second layer (middle arrow, Fig.8) of charge traps is $p_2=C\times\exp[-x_2/\xi]=C\times\exp[-2d_p/\xi]$, because the second layer is removed from the initial location of the tunneling electrons by distance $2d_p$. The third layer is removed by a distance $3d_p$, so the corresponding tunneling rate is (bottom arrow, Fig.8) $p_3=C\times\exp[-x_3/\xi]=C\times\exp[-3d_p/\xi]$, etc. The rate of tunneling events occurring over a distance of $n$ steps is $p_n=C\times\exp[-x_n/\xi]=C\times\exp[-nd_p/\xi]$. The underlying assumption here is that the relevant electronic states are localized, since they take place inside the dielectric layer.

Let us assume now that a voltage is applied and the electrons begin to tunnel into the dielectric or inside the dielectric, because the energy of the available traps is reduced and becomes comparable to the Fermi energy in the metallic electrode. Assume that we wait a time duration $t$ and then estimate how many charge-trap planes are charged (filled with electrons). It can be said that the time interval needed to achieve a tunneling of an electron by $n$ steps is $T_n\approx1/p_n$. So, within time $t$ all possible electron tunneling events will happen which are characterized by the condition $1/p_n<t$. The farthest charged penetration distance, $n_{max}(t)$, which will be filled within time $t$ is defined by the condition $T_n\approx t$, or $p_{n\_max}\approx1/t$, or $C\times\exp[-d_p n_{max}/\xi]\approx1/t$. This can be solved to obtain $n_{max}=(\xi/d_p)\times\ln(tC)$. So the charge which can enter the dielectric within time $t$ is $Q(t)=M_p\times n_{max}$, where $M_p$ is the number of traps per one trap plane. If the chosen number of planes on which the charges can shift is $N_p$, then $M_p=N_{trap}/N_p$, where $N_{trap}$ is the total number of charge traps in the sample. The number of layers is chosen for simplicity, so $N_p$ can be any integer number much larger than one. But $N_p$ should not be too large so that $M_p$ is still much larger than unity. In this model the distance between the planes is $d_p=d/N_p$, where $d$ is the total thickness of the dielectric.

We can now make a prediction for the total charge that enters the dielectric within time $t$, namely $Q(t)=M_p\times n_{max}=(\xi M_p/d_p)\times\ln(tC)$ (the model is only applicable if $t$ is sufficiently large so that the number of charged trap planes is larger than unity). Thus the charging current should depend on time as follows: $I(t)=(dQ/dt)=(\xi M_p/d_p)(1/t)$. This is very similar to the experimental observations.

The same current, presumably, exits the dielectric since tunneling is reversible if the voltage is removed. Thus the discharge current should also be defined as $I(t) = (\xi M_p / d_p)(1/t)$. Our experimental tests confirmed that the charging and discharging current curves are very similar (not shown). The power dependence of the experimental discharge curves is indeed similar to the predicted $1/t$ dependence (Fig.6).

## ACKNOWLEDGMENTS

This work was funded in part by the Office of Naval Research grant N00014-15-1-2397, Air Force Research Laboratory grant AF FA9453-14-1-0247 and NASA Contract NNX16CM29P. We thank Dr. Charles Marsh for fruitful discussions and the ARMY ERDC for supporting work with BPO W9132T-14-A -0001. This work was inspired by discussions at the Santa Fe Institute in Santa Fe, NM, USA.